# Low-cost approximate reconstructing of heterogeneous microstructures

W. Olchawa[a], R. Piasecki[a,*], R. Wiśniowski[a], D. Frączek[b],

[a] *Institute of Physics, University of Opole, Oleska 48, 45-052 Opole, Poland*

[b] *Department of Materials Physics, Opole University of Technology, Katowicka 48, 45-061 Opole, Poland*

HIGHLIGHTS

- A two-exponent power-law (TEPL) based on the entropic measure of spatial inhomogeneity is found.
- The model of overlapping spheres of a radius specified from the TEPL generates low-cost preferred configurations.
- The best of them provides an approximate reconstruction of a given three-dimensional medium.
- The proposed approach is tested on exemplary samples of ceramic and carbonate.
- The final reconstructions can serve as starting configurations.

ABSTRACT

We propose an approximate reconstruction of random heterogeneous microstructures using the two-exponent power-law (TEPL). This rule originates from the entropic descriptor (ED) that is a multi-scale measure of spatial inhomogeneity for a given microstructure. A digitized target sample is a cube of linear size $L$ in voxels. Then, a number of trial configurations can be generated by a model of overlapping spheres of a fixed radius, which are randomly distributed on a regular lattice. The TEPL describes the averaged maximum of the ED as a function of the phase concentration and the radius. Thus, it can be used to determine the radius. The suggested approach is tested on surrogate samples of ceramic and carbonate. In each of the cases, fifty low-cost trials provided a few good enough candidates to a selection of the optimal reconstruction. When a better accuracy is planned, the final reconstructions can serve as the starting configurations. Then, the resulting reconstructions should be competitive indeed to those starting with random configurations.

(Some figures in this article are in colour only in the electronic version)

*Keywords*: 3D microstructure reconstruction; Two-exponent power-law; Entropic descriptor; Heterogeneous microstructures

[*] Corresponding author.

*E-mail addresses*: wolch@uni.opole.pl (W. Olchawa), piaser@uni.opole.pl (R. Piasecki), ryszard.wisniowski@gmail.com (R. Wiśniowski), d.fraczek@po.opole.pl (D. Frączek).

## 1. Introduction

Undoubtedly, statistical reconstructing of three-dimensional (3D) random heterogeneous materials attracts much attention in computational materials science. The main point is to realize that some of the effective properties can be correlated with chosen spatial structural features. In other words, this connection can be termed a structure/property relationship [1]. On the other hand, 3D microstructure modelling allows fully capturing microstructural behaviour. It also makes easier the prediction of effective properties of disordered media [2]. Thus, an efficient method for statistical 3D reconstructing is an indispensable part of developing a reasonable model microstructure.

An additional challenge facing physicists who try to reconstruct heterogeneous media is how to preserve the geometric and other relationships of the phases of the given material. The outcomes allow successful rebuilding of the material with preserving the microstructure properties, even when the original construction process is unknown [3]. Additionally, such results facilitate optimal modification of macroscopic properties, materials design and characterisation across length scales [4–12]. The majority of reconstruction methods face the following two questions: "How to extract quantitative characteristics that represent the key microstructure features from digital images, and correspondingly, how to efficiently reconstruct statistically equivalent microstructures in the three-dimensional (3D) space for accurate structure–property relation assessments?" [13].

Motivated by the latter question, here the emphasis is put on the simplicity and computational efficiency of the proposed method. It is addressed to statistically homogeneous, isotropic porous materials. The basic idea of our approach is to use the so-called overlapping spheres model (OSM) with a single parameter, i.e., the sphere radius. Nevertheless, although the OSM is a flexible structural model, there are particular materials, which are hardly to be modelled in this way, e.g., particle reinforced composites. However, for large enough particle sizes and low filler concentrations, the method seems to be still applicable. On the other hand, fibre porous microstructures are definitely out of range of our method. Recently, an example of such porous microstructures have been discussed, namely biopolymer networks [14]. For the first time a variety of structural descriptors as well as the effective diffusive transport properties were calculated therein at different collagen concentrations.

The approach described in Section 2 can be included to a broad class of stochastic reconstructions based on various morphological descriptors with some recent modifications [15–26], to mention but a few. It is also worthy to notice a different branch of statistical reconstructions focused on nonstationary disordered materials and based on the so-called multiple-point statistics together with a cross-correlation function and one-dimensional raster path, see e.g., the recent paper [27] and citations therein.

Our method ensures approximate statistical similarity meant as the "distance" between the two curves. The first one relates to the so-called *entropic* descriptor (ED) [28, 29]. It can be computed for a digitized 3D target-image obtained, e.g., by X-ray computed tomography (CT) for real-world microstructure. The second ED-curve corresponds to the optimal low-cost reconstruction of the target. The basic definition of the ED provides Section 2 while the detailed description is given in Appendix A. To reconstruct an entire 3D medium the approach developed here does not make use of (i) the standard simulated annealing (SA), and (ii) a single 2D thin cross-section of a target. It should be stressed that our method can be applied to uncover what kind of a "synthetic" microstructure can be matched to a given "hypothetical" target ED-curve. In principle, for a given sample of a porous microstructure, one can readily provide a series of statistically similar model reconstructions. They can be then used for testing purposes.

We propose a splitting of the process of developing of the model microstructure into two steps. The reconstructing of a microstructure from a random configuration of voxels (keeping the phase concentrations) is rather slow. To accelerate it, instead of voxels we use the so-called overlapping spheres model (OSM) for the fast preparation of a set of approximate random 3D microstructures. Comparing the ED-curves, the 3D configuration of the optimal statistical similarity to the target microstructure can be then selected. If needed, the second step comes into play. Namely, the resulting approximate microstructure can be utilized again as the starting configuration by one of details oriented methods, e.g., with the usage of the SA. In this way, a more accurate reconstruction can be obtained with preserving some important geometrical characteristics, e.g., the value of an interface. Moreover, one can expect a substantial lowering of the computational cost of the overall reconstructing process. This work is devoted to the first step only.

It should be stressed that for some materials, the first step offers an acceptable approximation of its microstructure. In this context, one can point out those materials, whose effective properties are less sensitive to the structural details, like elastic moduli and effective conductivity [1]. On the other hand, one can identify effective properties of a heterogeneous material, which are length-scale dependent and thus, sensitive to the structural details. For example, the two important effective properties for the fluid-saturated porous media, i.e., trapping constant and scalar fluid permeability [30, 31] belong to this group. Then, the subsequent optimisation, which is the second step of the process of the modelling microstructure, is required.

The rest of the paper is organised as follows: Section 2 recalls the basic definition of the entropic descriptor serving as a measure of spatial inhomogeneity. The Section 2 provides also the uncovered two-exponent power-law and describes the related low-cost reconstructing procedure. In Section 3, the performance of the method is demonstrated with two illustrative examples of 3D reconstruction of the surrogate porous media. Finally, Section 4 contains concluding remarks.

## 2. The method

Let us consider a cube of the linear size $L$ composed of unit black/white voxels centred on sites of a regular lattice. The colour of a voxel can be attributed to a proper phase. In this way, various random two-phase microstructures are modelled by different spatial arrangements of the phase-corresponding voxels. The present approach is based on one of entropic descriptors (EDs) that describes the average multi-scale spatial inhomogeneity of a given microstructure [28, 29, 32, 33]. However, the estimation of the spatial inhomogeneity is related to the scale of its description. Therefore, for any length scale $1 \leq k \leq L$, the given cube is sampled by $\lambda(k)$ overlapping cells of size $k \times k \times k$. The chosen ED takes into account the statistical *dissimilarity* between the actual (current) macrostate AM($k$) with entropy $S(k) = \ln \Omega(k)$ and the most uniform reference (theoretical) one RM$_{max}$($k$) with maximal entropy, $S_{max}(k) = \ln \Omega_{max}(k)$. For convenience, Boltzmann's constant $k_B$ is set to unity. The $\Omega(k)$ and $\Omega_{max}(k)$ describe the numbers of microstates for realizations of the AM($k$) and RM$_{max}$($k$), respectively. Consequently, we use the difference of the corresponding entropies, $S_\Delta(k; 3D) \equiv [S_{max}(k) - S(k)]/\lambda(k)$, averaged per cell; see Appendix A. The $S_\Delta(k; 3D)$-function can be employed to quantitative characterisation of average spatial inhomogeneity for a target, as well as, trial microstructure.

Now, we would like to introduce an entropic descriptor based the two-exponent power-law (TEPL) for a random heterogeneous microstructure generated by means of the model of overlapping spheres. The uncovered formula can be written as

$$< \max S_\Delta(\phi, R; L) > = A(L) \phi^{0.41} R^{q(L)}, \qquad (1)$$

where $\log_{10} A(L) = 21.8/L + 0.37$, $q(L) = -45.5/L + 2.96$ and $L$ is a linear size of voxel-cube. The formula relates the arithmetic average of maximums of the spatial inhomogeneity denoted as $< \max S_\Delta(\phi, R; L) >$ to the variables $\phi$ and $R$. Here, $\phi$ means the volume fraction of matrix porous-phase (white voxels) called porosity, while $1 - \phi$ denotes the complementary fraction of a solid-phase (black voxels). In turn, $R$ is a radius of interpenetrating spheres of a black phase, which are randomly distributed on a regular lattice. Making use of the OSM, a number of random three-dimensional configurations can be easily generated. In our case, to prepare a statistical ensemble of a random variable, max $S_\Delta(\phi, R; L)$, one hundred random configurations were generated for every combination $\{\phi, R; L\}$ of the following variables: $\phi = 0.1, 0.2,..., 0.9$ and $R = 1.0, 1.5, …, 15$ with the chosen linear sizes $L = 80, 100, 130, 200, 300$. The obtained set of data allowed finding the final Formula (1). It is worth noticing the TEPL works well for middle-range phase fractions.

One remark is in order. For randomly generated configurations, we expect the following behaviour: the smaller radius $R$ is, the lower average spatial inhomogeneity should appear, so the $< \max S_\Delta(\phi, R; L) >$ should be lower, too. Such a behaviour can be observed if $q(L) > 0$ and consequently, Formula (1) can be used safely when $L > 15$. On the other hand, for larger linear sizes, i.e. for $L \to \infty$, the formula simplifies to the limiting form

$$< \max S_\Delta(\phi, R) > \cong 2.34 \, \phi^{0.41} R^{2.96} . \qquad (2)$$

At this stage, we are ready to present some details of the reconstructing procedure for a given heterogeneous microstructure. The key point is to obtain a number, say $N$, of low-cost but adequate trial three-dimensional configurations. To do it we employ the aforementioned model of overlapping solid-phase spheres of a *fixed* radius $R$. At the present stage, the value of $R$ is unknown. However, having calculated target's entropic descriptor, $S_\Delta(k; 3D, \text{target} = \mathbf{T})$, as a function of length scale $k$, we know the values of the max $S_\Delta(k; 3D, \mathbf{T})$ and the maximum-related length scale $k_{\max}(\mathbf{T})$. This allows temporary substituting in (1) the obtained max $S_\Delta(k; 3D, \mathbf{T})$ instead of the average value of the random variable, i.e. $< \max S_\Delta(\phi, R; L) >$. Now, for a given $\phi$ and $L$ the needed value of radius $R$ can be specified directly from (1). This way guarantees that for the generated current $N$-trial configurations, the simulated max $S_\Delta(\phi, R; L)$-values should be distributed around the value of max $S_\Delta(k; 3D, \mathbf{T})$. One can also expect that the estimated average value of the random variable should be very close to the target value, i.e. max $S_\Delta(k; 3D, \mathbf{T})$ in accordance with (1). Thus, any number of low-cost model configurations for given $\phi$ and $L$ can be obtained easily. All we have to do is to select among them a final configuration, for which the max $S_\Delta(\phi, R; L)$-value and the maximum related length scale $k_{\max}$ are the closest to their target counterparts, i.e. max $S_\Delta(k; 3D, \mathbf{T})$ and $k_{\max}(\mathbf{T})$.

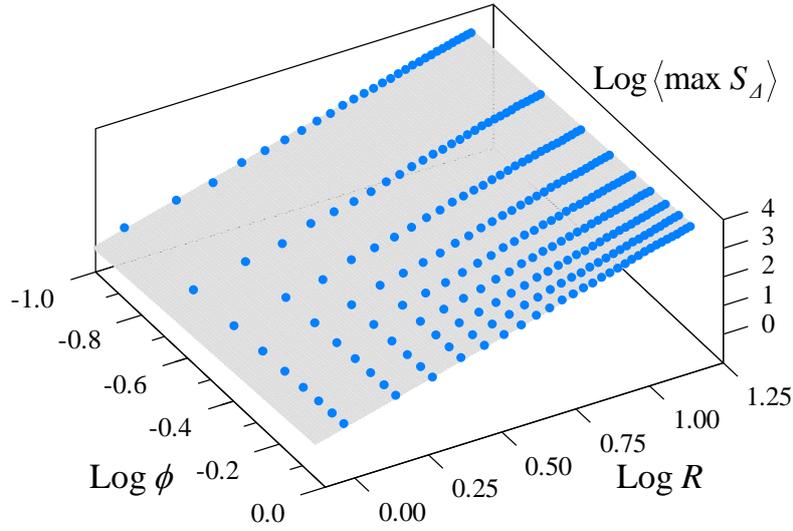

**Fig. 1.** A two-exponent power-law (TEPL) as a function of $\log_{10}(\phi)$ and $\log_{10}(R)$ tested for a fixed linear size $L = 160$ that differs from the earlier ones used in the uncovering of the TEPL. According to Formula (1), the theoretical plane (grey online) illustrates the predicted values of $\log_{10}(<\max S_\Delta(\phi, R; 160)>) \cong 0.51 + 0.41 \log_{10}(\phi) + 2.68 \log_{10}(R)$. For the additionally OSM-generated one hundred random configurations for each combination $\{\phi, R; L = 160\}$ of the earlier chosen variables $\phi = 0.1, 0.2, ..., 0.9$ and $R = 1.0, 1.5, ..., 15$, the "empirical" values of $\log_{10}(<\max S_\Delta(\phi, R; 160)>)$ are marked as filled circles (blue online). All the symbols show satisfactory locations on the plane. The best fitting appears for $5 < R < 10$. (For interpretation of the references to colour in this figure legend, the reader is referred to the web version of this article.)

Now, for test purposes the linear size parameter is fixed to $L = 160$ that differs from the values used earlier. Taking logarithm of both sides of Formula (1), the TEPL in action can be illustrated in Fig. 1. The resulting theoretical plane (grey online) is defined as $\log_{10}(<\max S_\Delta(\phi, R; 160)>) \cong 0.51 + 0.41 \log_{10}(\phi) + 2.68 \log_{10}(R)$. In order to make an independent examination of applicability of the TEPL, additional one hundred random configurations are generated for each combination $\{\phi, R; L = 160\}$ with the earlier selected values of variables $\phi$ and $R$. In Fig. 1, the corresponding "empirical" values of $\log_{10}(<\max S_\Delta(\phi, R; 160)>)$ are marked as filled circles (blue online). As expected, one can observe the suitable placement of the symbols on the theoretical plane, but the best fitting appears for $5 < R < 10$. In the next section, we apply our method to ceramics and carbonate samples having different 3D microstructures.

## 3. The illustrative examples

Here we show how our method works on two examples of surrogate porous medium, where voids are treated as the "white" phase. The entropic descriptor based reliable 3D reconstructions were obtained earlier from a corresponding single 2D input image of the real porous ceramics and carbonate samples. The respective porosities are $\phi = 0.3814$ and $0.1438$ [34]. The reconstructed microstructures represented by voxel-cubes of linear size $L = 300$ are treated here as two digitized targets. For each of the final 3D configurations, any of the cross-sections perpendicular to the main axes is statistically similar to the appropriate input 2D image. The related statistical "distance" was measured by the entropic cost function per plane. The averaged objective function was the sum of squared and normalized differences between the values of normalized EDs related to a current plane configuration and the target pattern [34].

Here, the given target values, i.e. max $S_\Delta(k; 3D, \mathbf{T}) = 736.78$ (for ceramics) and $1428.43$ (for carbonate) appear at length scales $k_{\max}(\mathbf{T}) = 45$ and $64$, cf. Figs. 2a and 3a with the thick solid curve (red online), respectively. Further, we determine the needed values of the radius $R$ from the two-exponent power-law given by Formula (1) as it was described in the previous section. Correspondingly, they are $R \cong 8.36$ and $12.20$. Now, for each of the two considered exemplary microstructures, we generate $N = 50$ trial 3D configurations, making use of the OSM. In this way, the computed $S_\Delta(k; 3D)$-curves show the maximums well spread around the related target counterparts. However, it should be stressed that their locations at $k_{\max}(i)$, $i = 1,\ldots,N$, are widely spread in the nearby proper scale $k_{\max}(\mathbf{T})$. The reason is that this quantity is difficult to control in the present simple approach. Then, among the curves of each of the two sets, we select a few preliminary curves with the $i$th values of max $S_\Delta(\phi, R; 300)$ and $k_{\max}(i)$ sufficiently close to the corresponding target counterparts. The final choice of an optimal curve is based on the best qualitative similarity with the target one, especially at length scales ranged around the main peak of $S_\Delta$.

For the purpose of comparison, the target $S_\Delta(k; 3D, \mathbf{T})$ thick solid curve (red online) and selected $S_\Delta(k; 3D)$ thin solid and dashed curves are collected in Fig. 2a (3a), for the ceramics (carbonate) samples, respectively. The thin solid curve (blue online) relates to the trial optimal-microstructure, while the thin dashed lines describe the selected other reconstructions. Correspondingly, in Fig. 2b (3b) a 3D exterior view of the target-cube for the ceramics (carbonate) sample is shown. Similarly, in Fig. 2c (3c) a 3D exterior view of the corresponding optimal reconstruction for the ceramics (carbonate) sample is displayed. We point again that the TEPL acts properly for middle-range phase fractions. On the other hand, the given target should be a representative sample of the real material.

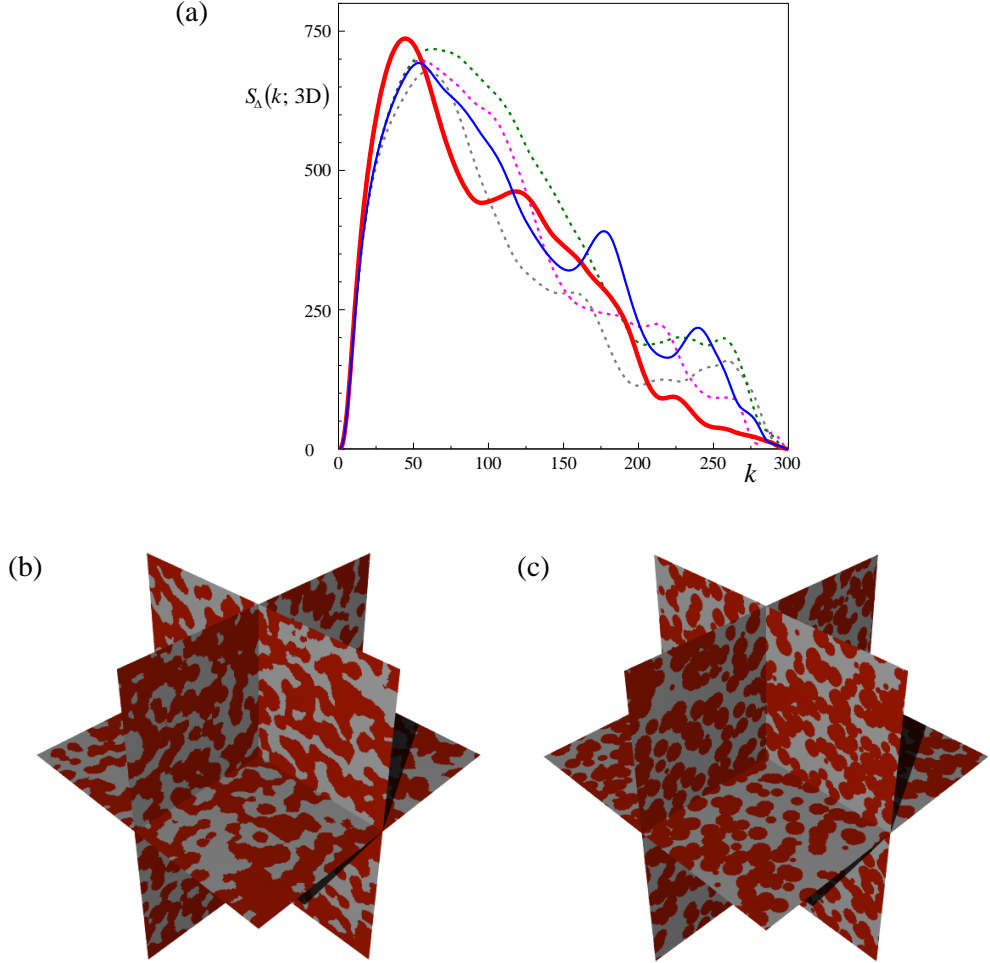

**Fig. 2.** The illustration of the TEPL-method of the low-cost approximate reconstructing of heterogeneous microstructures. It is based on the chosen entropic descriptor quantifying a multi-scale spatial inhomogeneity. A surrogate sample of ceramic is a target-cube of linear size $L = 300$ with a given porosity $\phi = 0.3814$. (a) The thick solid curve (red online) describes the entropic descriptor $S_\Delta(k; 3D, \mathbf{T})$ for the target ceramic microstructure, correspondingly the thin solid curve (blue online) for its optimal reconstruction while the thin dashed lines relate to other reconstructions selected as preliminary ones among $N = 50$ trials. (b) A cross section of the target-cube. (c) A cross section of its optimal reconstruction. (For interpretation of the references to colour in this figure legend, the reader is referred to the web version of this article.)

One can expect that a similar TEPL could be found for the so-called statistical complexity measure $C_\lambda$ discussed in Ref. [35]. More precisely, it can be written shortly as $C_\lambda(k; 3D) = S_\Delta(k; 3D)\,\gamma(k)$, where $\gamma(k) \equiv [S(k) - S_{\min}(k)]/[S_{\max}(k) - S_{\min}(k)]$. An evident inequality $0 \leq \gamma(k) \leq 1$ always holds. Notice, that at smaller length scales the significance of the component $S_{\min}(k) = \ln \Omega_{\min}(k)$ increases. This term describes the lowest possible value of the current entropy $S(k)$. The minimum is accessible for the most spatially non-uniform reference macrostate. Thus, additional spatial features of the given microstructure could possibly be accounted for.

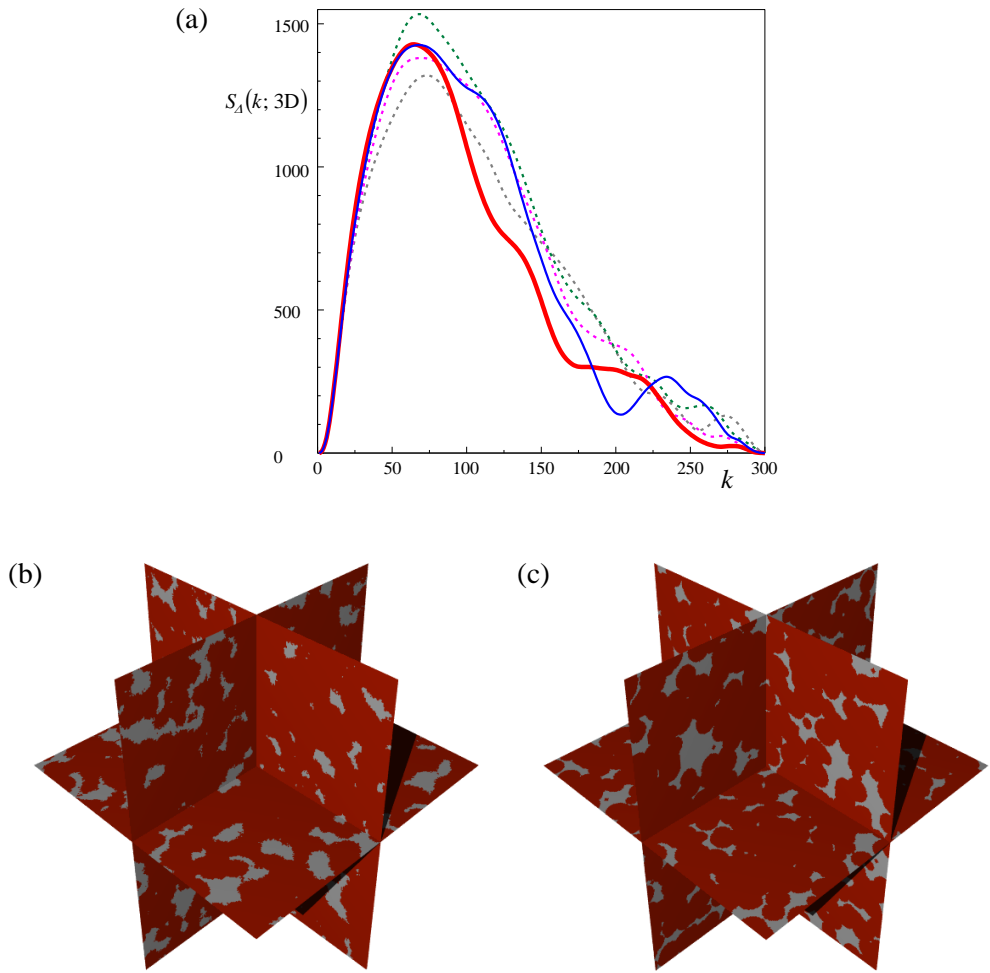

**Fig. 3.** The same as in Fig. 2 but for a surrogate sample of carbonate with a given porosity $\phi = 0.1438$.

We note that, if desired, one could potentially extend the used approach to three-phase materials via the so-called decomposable multiphase entropic descriptor [36]. All we need is to find a phase related power-law formula being a counterpart of the TEPL given by (1). For example, such multi-phase microstructures can be found in Ref. [37]. However, the overlapping sphere model used therein was defined in a different context.

## 4. Conclusions

We propose a simple method for low-cost reconstructing heterogeneous microstructures. It employs the uncovered two-exponent power-law that originates from the chosen entropic descriptor. This descriptor quantifies an averaged multiscale spatial inhomogeneity. The basic ingredient of our method is the generation of trial configurations by randomly distributed overlapping solid-phase spheres of a fixed radius. For a digitized target microstructure, this radius results from the two-exponent power-law formula. Our approach is computationally a very efficient one. However, it provides approximate statistical reconstructions. When tested on surrogate porous samples of ceramic and carbonate, the preliminary selection among fifty low-cost trials offers a few candidates to the final selection of the optimal reconstruction. The optimal reconstructions can be useful as three-dimensional starting configurations for the improving accuracy if needed. Actually, not only the average degree of non-uniformity, but additional spatial features important from the effective properties viewpoint can also be incorporated.

**Appendix A**

In general, the given binary cube of size $L \times L \times L$ can be sampled by $\lambda(k) = ((L-k)/z + 1)^3$ overlapping 3D-cells of size $k \times k \times k$ with a sliding factor $1 \leq z < k$ in three main directions provided $((L-k) \bmod z) = 0$. Here, the chosen $z = 1$ gives the maximal non-trivial overlapping of the cells. So, at each of the considered scales $k$, we analyse the so-called auxiliary 3D-map $L_a(k) \times L_a(k) \times L_a(k)$, where $L_a(k) \equiv (L-k+1)\,k$. The maps are composed of the sampled 3D-cells placed in a non-overlapping manner. They can be treated as the representative and length scale dependent instances of the investigated microstructure since in this way general spatial features are clearly reproduced; for a grey-level pattern *cf*. Fig. 1b in Ref. [38]. Keeping this in mind, the basic constraint at every length scale $k$ for the cell occupation numbers $n_i(k; 3D)$ of unit voxels of a given phase reads

$$\sum_{i=1}^{\lambda} n_i(k;3D) = N(k;3D). \tag{A1}$$

Here, $N(k; 3D)$ stands for the total number of unit voxels of a given phase for auxiliary 3D-map created at scale $k$. The non-trivial volume concentration $\phi(k) \equiv N(k;3D)/\lambda(k)\,k^3$ is assumed, that is $0 < \phi(k) < 1$. Very small fluctuations of $\phi(k)$ at different length scales can be discarded within our approach. To simplify the notation we will omit the parameter $k$ and symbol "3D" wherever it does not lead to misunderstanding.

Then, at each scale $k$, we consider two appropriate configurations for every 3D-map. The first arrangement corresponds to the real structure for which the actual micro-canonical entropy, $S(k) = \ln \Omega(k)$, is calculated. The second one is the most uniform theoretical configuration with the occupation numbers given below, which ensure the highest possible value of the reference entropy, $S_{max}(k) = \ln \Omega_{max}(k)$. Then, the chosen entropic descriptor for the average multi-scale spatial inhomogeneity of a given microstructure has the following form

$$S_\Delta(k; 3\text{D}) = [S_{max}(k) - S(k)]/\lambda(k) \ . \tag{A2}$$

The numbers of realizations of appropriate macrostates are listed below. For the actual macrostate $\text{AM}(k) \equiv \{n_i(k)\}$, $i = 1, 2,\ldots, \lambda(k)$, the clear number $\Omega(k)$ of its realizations is given by the product of the ways that each of the sampled cells can be occupied with the number $n_i$ of the phase voxels under the above constraint (A1)

$$\Omega(k) = \prod_{i=1}^{\lambda} \binom{k^3}{n_i} . \tag{A3}$$

In turn, the maximum possible value $S_{max}(k)$ is accessible for the most spatially homogeneous (at this scale $k$) reference macrostate. This macrostate can be written in a detailed form as follows: $\text{RM}_{max}(k) \equiv \{(\lambda - r_0)\, n_0;\, r_0\, (n_0 + 1)\}_{max}$, where exactly $(\lambda - r_0)$ and $r_0$ of the sampled cells is occupied by the $n_0$ and $n_0 + 1$ of the phase voxels, respectively. Thus, the simple relation holds: $N(k) = (\lambda - r_0)\, n_0 + r_0\, (n_0 + 1) \equiv \lambda\, n_0 + r_0$, where $r_0 = (N \bmod \lambda)$, $r_0 \in (0, 1, ..., \lambda - 1)$ and $n_0 = (N - r_0)/\lambda$. Now, the number of proper microstates reads

$$\Omega_{max}(k) = \binom{k^3}{n_0}^{\lambda - r_0} \binom{k^3}{n_0 + 1}^{r_0} . \tag{A4}$$

The first maximum of $S_\Delta(k; 3\text{D})$, observed at the characteristic scale $k_{max}$, quantifies a maximal spatial inhomogeneity per cell. Such a situation indicates formation of clusters of a characteristic range of sizes, which are comparable with $k_{max}$. We recommend the additional and more instructive numerical example presented in Appendix A of Ref. [29].

## References

[1] S. Torquato, Random Heterogeneous Materials, Springer, New York, 2002.
[2] M. Sahimi, Heterogeneous Materials I & II, Springer, New York, 2003.
[3] M. A. Davis, S. D. C. Walsh, M. O. Saar, Phys. Rev. E 83 (2011) 026706.
[4] S. Torquato, Annu. Rev. Mater. Res. 32 (2002) 77.
[5] S. Torquato, Microstructure optimization, in: Y. Sidney (Ed.), Handbook of Materials Modeling, Springer-Verlag, New York, 2005.
[6] S. Torquato, Annu. Rev. Mater. Res. 40 (2010) 101.

[7] D. Fullwood, S. Niezgoda, B. Adams, S. Kalidindi, Prog. Mater. Sci. 55 (2010) 477.
[8] B.L. Adams, S.R. Kalidindi, D.T. Fullwood, Microstructure-Sensitive Design for Performance Optimization, Butterworth-Heinemann, 2012.
[9] Y. Liu, M.S. Greene, W. Chen, D.A. Dikin, W.K. Liu, Comput. Aided Des. 45 (2013) 65.
[10] D.D. Gerrard, D.T. Fullwood, D.M. Halverson, Comput. Mater. Sci. 91 (2014) 20.
[11] D.S. Li, JOM 66 (2014) 444.
[12] R. Dingreville1, R.A. Karnesky, G. Puel, J-H. Schmitt, J. Mater. Sci. 51 (2016) 1178.
[13] H. Xu, D.A. Dikin, C. Burkhart, W. Chen, Comput. Mater. Sci. 85 (2014) 206.
[14] Y. Jiao, S.Torquato, Phys. Biol. 9 (2012) 036009.
[15] Y. Jiao, F.H. Stillinger, S.Torquato, Proc. Natl Acad. Sci. USA 106 (2009) 17634.
[16] P. Čapek, V. Hejtmánek, J. Kolafa, L. Brabec, Transp. Porous Media 88 (2011) 87.
[17] Y. Jiao, E. Padilla, N. Chawla, Acta Mater. 61 (2013) 3370.
[18] C.E. Zachary, S. Torquato, Phys. Rev. E 84 (2011) 056102.
[19] E. Guo, N. Chawla, T. Jing, S. Torquato, Y. Jiao, Mater. Charact. 89 (2014) 33.
[20] L.M. Pant, S.K. Mitra, M. Secanell, Phys. Rev. E 90 (2014) 023306.
[21] Y. Jiao, N. Chawla, Integr. Mater. Manuf. Innov. 3 (2014) 3.
[22] D.D. Chen, Q. Teng, X. He, Z. Xu, Z. Li, Phys. Rev. E 89 (2014) 013305.
[23] S. Chen, H. Li, Y. Jiao, Phys. Rev. E 92 (2015) 023301.
[24] M.L. Gao, X.H. He, Q.Z. Teng, C. Zuo, D.D. Chen, Phys. Rev. E 91 (2015) 013308.
[25] R. Bostanabad, A.T. Bui, W. Xie, D.W. Apley, W. Chen, Acta Mater. 103 (2016) 89.
[26] D.M. Turner, S.R. Kalidindi, Acta Mater. 102 (2016) 136.
[27] P. Tahmasebi, M. Sahimi, Phys. Rev. E 91 (2015) 032401.
[28] R. Piasecki, Physica A 277 (2000) 157.
[29] R. Piasecki, Proc. R. Soc. A 467 (2011) 806.
[30] S. Torquato, D.C. Pham, Phys. Rev. Lett. 92 (2004) 25505.
[31] D.C. Pham, S. Torquato, J. Appl. Phys. 97 (2005) 013535.
[32] R. Piasecki, W. Olchawa, Modell. Simul. Mater. Sci. Eng. 20 (2012) 055003.
[33] W. Olchawa, R. Piasecki, Comput. Mater. Sci. 98 (2015) 390.
[34] D. Frączek, W. Olchawa, R. Piasecki, R. Wiśniowski, arXiv:1508.03857v2 [cond-mat.stat-mech].
[35] R. Piasecki, A. Plastino, Physica A 389 (2010) 397.
[36] D. Frączek, R. Piasecki, Physica A 399 (2014) 75.
[37] E.J. Garboczi, Physica A 442 (2016) 156.
[38] R. Piasecki, Physica A 388 (2009) 4229.